\documentclass[11pt]{article}
\usepackage{geometry}
\usepackage[parfill]{parskip}\usepackage{graphicx}
\usepackage{amssymb}
\usepackage{epstopdf}
\makeatletter
\def\blfootnote{\xdef\@thefnmark{}\@footnotetext}
\makeatother
\def\struct#1{{\bigskip\par\noindent \sc #1 \nopagebreak[4]\par\noindent\nopagebreak[4]}}

\title{A Note on Approximate Nearest Neighbor Methods}
\author{Thomas Breuel}
\date{}
\begin{document}
\maketitle
\blfootnote{
\hskip-4ex This technical report was originally written in September 2005 and had limited distribution by E-mail at the time.
}

\section{Background}

\struct{The Nearest Neighbor Problem}

The problem of finding nearest neighbors to a query point in a database has a long history in computer science.
Formally, we can state the problem as follows.
Assume we are given a space $V$ of samples or objects, together with a function $d: V \times V \rightarrow \mathbb{R}$.
Commonly, $V$ will be a vector space and $d$ is defined by a metric in that vector space; 
that is the case we will be considering in this paper.

The nearest neighbor problem is the problem of finding, given a finite set, the database, $S = \{s_1,\ldots,s_N\} \subseteq V$ and a query point $q \in V$, a distance $\rho$, and a set of results $R\subseteq S$, such that for all $r\in R$, $d(r,q)=\rho$, and for all $s\in S$, $d(s,q)\geq \rho$.

Many applications only require a single random representative from the result set $R$ (see the discussion below).
For non-singular sample distributions, this is, for practical purposes, equivalent to choosing an arbitrary representative from $R$ (since $\#R=1$ with probability $1$ for such distributions).

The references (e.g., Indyk and Motwani, 1999, Datar et al., 2004) contain extensive discussions of the complexity and applications of nearest neighbor algorithms and the motivations for introducing approximations.
Complexity does not directly concern us in this paper, so we will not discuss it further; this paper is concerned with the utility and implications of the approximation scheme chosen by recent work on fast, approximate nearest neighbor algorithms.

\struct{Nearest Neighbor Classification}

Nearest neighbor classification uses nearest neighbor algorithms to solve a classification task.
The classification task is defined by having a training set $T = \{(\omega_1,v_1),\ldots,(\omega_N,v_N)\}$ of pairs of a class $\omega$ and a sample vector $v$ each.
The training set is assumed to have been drawn independently from a joint density $p(\omega,v)$.
The task of a classifier is to estimate the class $\omega$ given just $v$ for a novel sample $(\omega,v)$ drawn according to $p(\omega,v)$.

There are many possible approaches to this problem.
The Bayes optimal answer is to classify sample $v$ as $\arg\max_{\omega} P(\omega|v)$, where $P(\omega|v)$ is the posterior probability derived from $P(\omega,v)$.
Of course, $P(\omega|v)$ is not given, so it needs to be estimated from the training set $T$.

Nearest neighbor methods provide one of the simplest and most robust methods, and do not even require explicit estimation of $P(\omega|v)$.
The nearest neighbor approach to classification says that if $v_i$ is the closest vector in the training set $T$ to the query point $v$, then we should classify the query point $v$ as being in class $\omega_i$ (the class associated with $v_i$ in the training set).

Nearest neighbor classification is easily seen to yield an error rate that is asymptotically within a factor of two of the Bayes optimal rate.

\struct{Approximate Nearest Neighbor Algorithms}

The literature on approximate nearest neighbors examines the complexity of exact nearest neighbors algorithms carefully and concludes that it is undesirably high (e.g., Indyk and Motwani, 1999); 
in essence, exact nearest neighbor algorithms do not yield significant speedups compared to brute force linear search, and they often require exponential space for the results of preprocessing.
In computer science, it is common to consider algorithms for computing approximate solutions when the computational complexity of exact solutions is undesirably high.
The literature defines an $\epsilon$-approximate nearest neighbor as follows.
If the nearest neighbor $v$ to some query point $q$ has distance $s$, then any vector $v'$ such that $d(v',q)\leq (1+\epsilon)s$ is an $\epsilon$-approximate nearest neighbor.

\section{Limitations of Approximate Nearest Neighbor Algorithms}

\struct{Approximations and Costs}

The utility of such approximations relies on assumptions about the relationship between the cost of making an approximation relative to the original solution.
For example, if the database represents travel destinations and the query point represents the current location of an airplane, then then cost of traveling to an $\epsilon$-approximate nearest neighbor from a query point in a straight line is at most higher by a factor of $(1+\epsilon)$ compared to the optimal solution.

However, the relationship between approximation and ``cost'' of a solution need not be linear.
For example, the cost of picking an $\epsilon$-approximate neighbor could be proportional not to the difference in distances between the optimal answer and the approximation, but to the volume of the shell between the two, that is, as $(1+\epsilon)^{r-1}$, where $r$ is the dimension of the space.

\struct{Approximate Nearest Neighbors}

A first indication that approximate nearest neighbors may not be very useful in high dimensions can be inferred simply by looking at the kinds of distance distributions we expect when querying a database with a data point.  

It is a well known (and frequently rediscovered) fact that in high-dimensional spaces, everything is about equally close to everything for many point distributions (see Aggarwal et al., 2004, Beyer et al., 1999).

While this can be derived formally from the central limit theorem, we can illustrate this easily with a simple computational experiment.

\begin{figure}[h]
\begin{center}
\includegraphics[height=10cm]{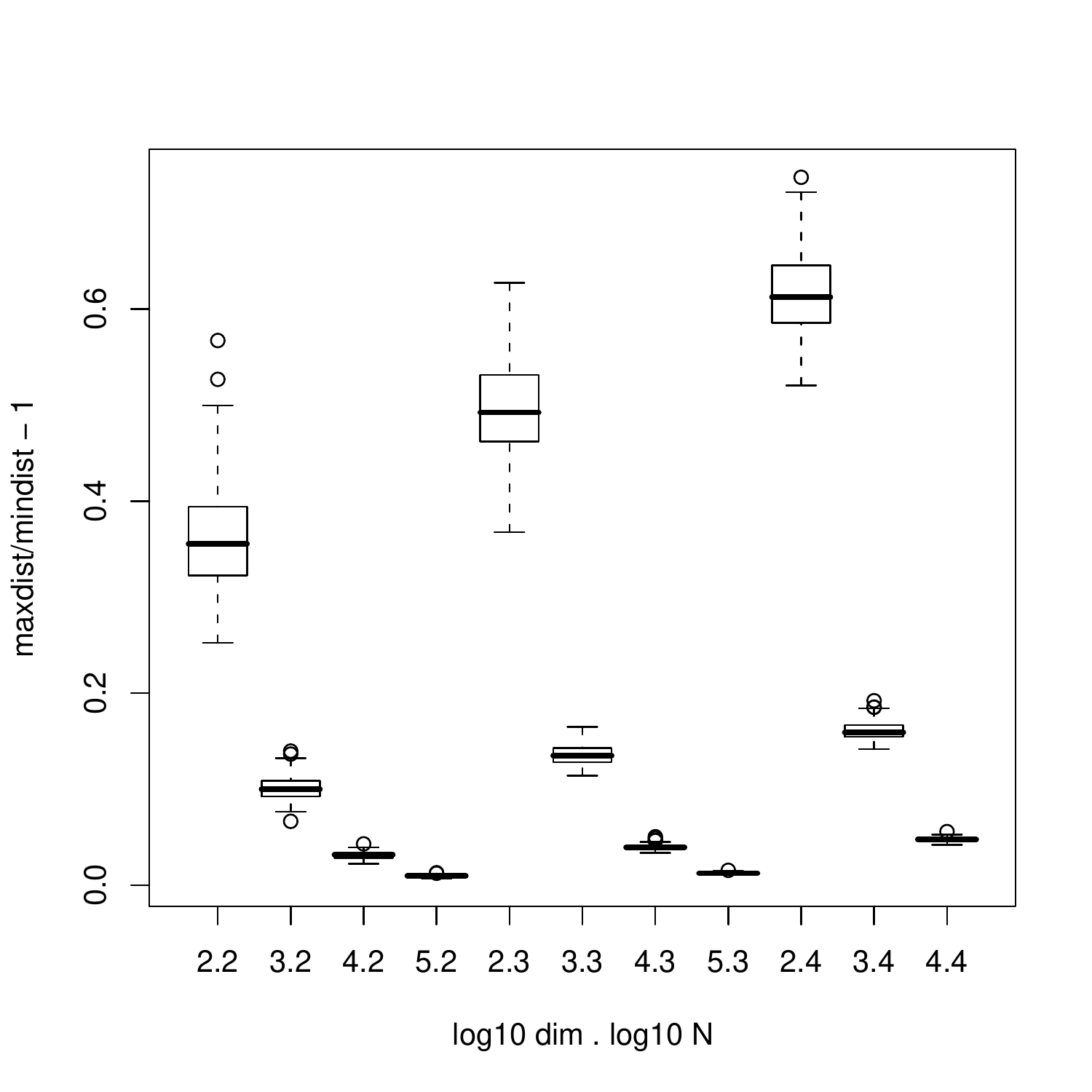}
\caption{Ratio of extreme values of the distance of points in a database from a query point.  The figures show that, for example, at a dimension of 10000 and $\epsilon=0.1$, returning an arbitrary point from the database will be an $\epsilon$-approximate nearest neighbor with high probability.  Each box in the boxplot corresponds to a 100 sample run.}
\label{extreme-values}
\end{center}
\end{figure}

For the experiment, we generate a database of 100, 1000, 10000, and 100000 vectors of dimension 100, 1000, 10000, and 100000, together with a query point, by drawing from a normal distribution with mean zero and standard deviation one.  We then measure the ratio of the furthest neighbor to the nearest neighbor.  Obviously, if $1+\epsilon$ is greater than this value, then any point whatsoever in the database is an $\epsilon$-approximate neighbor.  We see that for large dimensions, even small values of $\epsilon$ include the entire database as $\epsilon$-approximate neighbors. So, ultimately, as the dimensionality of the problem gets larger, approximate nearest neighbor methods can degenerate into trivially picking any data point arbitrarily.

This observation does not mean that specific approximate nearest neighbor methods are useless; in fact, they may well return far better results with high probability.  But what it shows is that analyzing the worst-case asymptotic complexity of $\epsilon$-approximate algorithms is meaningless.

\struct{Nearest Neighbor Methods on Low-Dimensional Surface}

It has been suggested that $\epsilon$-approximate nearest neighbor methods perform particularly well on problems in which data is distributed non-uniformly in a high-dimensional space.
A case of particular interest is where data is known to lie on or near a low-dimensional surface.
Good performance of approximate nearest neighbor methods in such cases has, in fact, been demonstrated.

However, that fact is not particularly surprising: many nearest neighbor algorithms perform well when then data lies on low-dimensional subspaces.
Work on approximate nearest neighbor methods have failed to demonstrate that such algorithms have an advantage in such cases compared to traditional exact nearest neighbor methods.

\struct{Approximations and Loss of Randomness}

There is a second problem with the definition of the $\epsilon$-approximate nearest neighbor problem.
In particular, it permits the algorithm to return an arbitrary sample that satisfies the approximation condition.

However, the proof of the asymptotic properties of nearest neighbor classifiers relies crucially on the fact that nearest neighbor samples are unbiased samples from the local distribution around the query point (Duda, Hart and Stork, 2003).
In essence, the proof considers the limit in which enough samples have been drawn from the sample distribution so that the nearest neighbor $s$ is sufficiently close to the query point $q$ that $P(\omega|s) \rightarrow P(\omega|q)$; we will be using the same limit in all the considerations below.

The probability of error is then seen to be simply the probability that two classes drawn at random from a binomial (or multinomial) distribution have different classes.
In the two-class case, the probability of error is $(1-E)E + E(1-E) = 2E-E^2 \leq 2E$, where $E = \min(P(1|q),P(2|q))$, the Bayes-optimal error rate.

But if $\epsilon$ is sufficiently large that multiple neighbors satisfy the $\epsilon$-approximation condition, and that case necessarily occurs for $\epsilon$-approximate nearest neighbor algorithms to represent a useful tool, according to its definition, the approximation algorithm is free to choose an arbitrary approximate nearest neighbor.

For example, the approximate nearest neighbor algorithm could consistently pick representatives of the least frequent class near a query point whenever such a representative is available.
Let $H$ be the probability that the neighborhood of the query point contains a member of the less frequent class.
By considering the different cases (class of the query point and existence of a neighbor in the less frequent class), we see that the error rate is $H(1-E)(1-E)+H(1-E)E+(1-H)E$.
We see that $H \approx 1-(1-E)^{k_\epsilon}$, where $k_\epsilon$ is the number of neighbors in the $\epsilon$-approximation, a number that rapidly approaches $1$ as $k_\epsilon$ grows.
If we set $H\approx 1$, then the error rate is approximately $1-E-2E^2$.

An even worse case appears if the approximation algorithm consistently picks approximate nearest neighbors that are in the wrong class; in that case, the error rate is simply the probability that a neighbor of the wrong class exists among the $k_\epsilon$ neighbors, that is, it is $E(1-(1-E)^{k_\epsilon}) + (1-E)(1-E^{k_\epsilon})$, which is seen to approach $1$ rapidly as $k_\epsilon$ grows.

Note that these worse bounds hold true even if $k_\epsilon$ is bounded.
Furthermore, for some distributions, it is not necessary for the approximation algorithm to be adversarial or even have knowledge of the classes to make consistently bad choices.

\struct{Singular Distributions}

To illustrate these points, consider the case of singular probability distributions over a vector space.  
Examples of such distributions are discrete distributions (distributions that are entirely mixtures of delta functions), mixtures of delta functions and smooth densities, and densities with a non-zero probability of returning the same sample multiple times.

Above, we saw that the condition that approximate nearest neighbors permit arbitrary representatives to be returned implies that such algorithms cannot guarantee the asymptotic bounds on classification error rate. 
The same analysis applies to exact nearest neighbor algorithms returning arbitrary representatives from the set of nearest neighbors when used with distributions that have a non-zero probability of returning the same sample point multiple times.
(Note that this case does occur in practice and can significantly affect the performance of nearest neighbor methods.)

A simple idea for transforming an exact nearest neighbor algorithm returning an arbitrary sample into one returning a random sample is to perturb each point in the database by a small amount;
for example, for a discrete distribution on the integer grid, perturbing the position of each sample and query point by less than $\frac{1}{4}$ is sufficient to ensure return of an exact but random nearest neighbor even using a nearest neighbor algorithm that is only guaranteed to return an arbitrary nearest neighbor (treatment of perturbations for the general case is more complex).
But the same approach fails for obtaining random representatives from an approximate nearest neighbor algorithm.
This example suggests that returning non-random representatives may be an essential feature of approximate nearest neighbor algorithms.

\section{Discussion}

\struct{Summary of Problems with Approximate NN}

This note has pointed out two problems with the definition of $\epsilon$-approximate nearest neighbors:
\begin{itemize}
\item for even simple distributions and large enough dimensions, a randomly chosen point becomes an $\epsilon$-approximate neighbor with high probability
\item establishing that an algorithm is an $\epsilon$-approximate nearest neighbor algorithm results in no guarantees about its performance in many common uses of nearest neighbor algorithms
\end{itemize}
The fundamental problem lies not with the algorithms, but with the notion of ``approximation'' used by $\epsilon$-approximate nearest neighbor methods; the implicit assumption that a close approximation leads to only a small increase in the cost of a solution is not justifiable in the context of nearest neighbors (in fact, a formal treatment of ``cost'' might show that the cost of an $\epsilon$-approximate nearest neighbor grows as $(1+\epsilon)^(r-1)$, where $r$ is the dimensionality of the space).
Furthermore, as we have seen, the approximation scheme neglects important requirements of common uses of nearest neighbor methods, such as returning a random sample, not just an arbitrary sample; this objection applies even in cases where an approximate nearest neighbor algorithm actually solves the problem as stated and the number of actual approximate nearest neighbors is small.
In different words, the ``curse of dimensionality'' has not been removed, it has merely been hidden in the approximation scheme.

\struct{Implications}

These results do not necessarily mean that the $\epsilon$-approximate nearest neighbor algorithms proposed in the literature are useless in practice.
What it means is that establishing that an algorithm is asymptotically an $\epsilon$-approximate nearest neighbor algorithm gives us no useful guarantees or predictions about the behavior or utility of that algorithm when used for the kinds of purposes that nearest neighbor algorithms are usually used for: it neither has useful meaning asymptotically (as the dimension grows), nor does it make useful predictions about its behavior on practical problem instance.
Whether the proposed algorithms are useful depends on other properties; maybe theoretical notions of nearest neighbor approximation can be defined that make both useful predictions and can be established for these algorithms.

Since the theoretical analyses are not useful, all non-exact nearest neighbor algorithms should be viewed as heuristic for the time being.
It may well turn out that some of those algorithms are useful and more efficient for specific problems or even for general classes of distributions, but this remains to be determined empirically and experimentally on standardized datasets.

\struct{Alternative Definitions}

While the notion of approximation underlying $\epsilon$-approximate nearest neighbor algorithms is problematic, there are some potential alternatives.
First, an approximation algorithm that returns a nearest neighbor that is within a given fractile $\epsilon$ of the distribution of nearest neighbors would likely be a better choice; in fact, we might want $\epsilon$ to be a strictly decreasing function of the size of the database.
Second, in order to address the concerns about randomness, we might impose at a minimum that the algorithm makes guarantees that approximate query results returned by the algorithm are (at least asymptotically, as the size of the database increases) representative of the class conditional densities around the query point.
We note as an aside that the $O(\log n)$ result in Datar et al. (2004) appears to reproduce well-known results on the complexity of nearest neighbor search in ultrametric spaces.

\section*{References}

{\parindent0pt
C. Aggarwal et al. (2001): On the surprising behavior of distance metrics in high dimensional space

K. Beyer et al. (1999): When is Nearest Neighbor meaningful.

T. Liu et al. (2004): An investigation of practical approximate nearest neighbor algorithms.  NIPS.

P. Indyk and R. Motwani (1999): Approximate nearest neighbors: towards removing the curse of dimensionality.

M. Datar et al. (2004): Locality-sensitive hashing scheme based on p-stable distributions.

R. O. Duda, P. E. Hart, and D. G. Stork (2003): Pattern Classification.
}

\end{document}